%% file: main.tex
\documentclass{IEEEtran}

\usepackage[table,xcdraw]{xcolor}
\usepackage{float}
\usepackage{multirow}
\usepackage{booktabs}
\usepackage{adjustbox}

\begin{document}


\title{Performance/power assessment of CNN packages\\on embedded automotive platforms}

\author{
Paolo Burgio, Gianluca Brilli\\
 {University of Modena and Reggio Emilia, Italy}\\
 {\{firstname.lastname\}@unimore.it}
}
%

\maketitle

\begin{abstract}
The rise of power-efficient embedded computers based on highly-parallel accelerators opens a number of opportunities and challenges for researchers and engineers, and paved the way to the era of \textit{edge computing}.
At the same time, advances in embedded AI for object detection and categorization such as YOLO, GoogleNet and AlexNet reached an unprecedented level of accuracy (mean-Average Precision -- mAP) and performance (Frames-Per-Second --FPS).
Today, edge computers based on heterogeneous many-core systems are a predominant choice to deploy such systems in industry 4.0, wearable devices, and --our focus-- autonomous driving systems.
In these latter systems, engineers struggle to make reduced automotive power and size budgets co-exist with the accuracy and performance targets requested by autonomous driving.
We aim at validating the effectiveness and efficiency of most recent networks on state-of-the-art platforms with embedded commercial-off-the-shelf System-on-Chips, such as Xavier AGX, Tegra X2 and Nano for NVIDIA and XCZU9EG and XCZU3EG of the Zynq UltraScale+ family, for the Xilinx counterpart.
Our work aims at supporting engineers in choosing the most appropriate CNN package and computing system for their designs, and deriving guidelines for adequately sizing their systems.
\end{abstract}


\input{introduction}
\input{related_works-full}
\input{architectures}
\input{neural_networks}

\input{experiments-hpcs}
\input{experiments_2}

\input{conclusions}

\nocite{*}

\bibliographystyle{abbrv}
\bibliography{biblio}

\clearpage

\appendix
\input{appendix_methodology}
\input{appendix_frameworks}

\end{document}

%% file: introduction.tex
\section{Introduction}
\label{sec:introduction}
The next generation of automotive systems will be powered by high-performance embedded computers, that exhibit TFLOPs of performance within a reduced Size, Weight and Power consumption (SWaP)~\cite{xavier, drive_px, versal, mobileye, ultrascale-zcu102}.
There is a huge interest from car-makers, but also from Tier1s, to use those platforms as a baseline for next-generation Advanced Driving Assistance Systems (ADAS) and Autonomous Vehicles (AV).
Notable players are Tesla, BMW, FCA, and Toyota, but also companies that traditionally did not belong to the automotive domain such as NVIDIA~\cite{drive_px} and Intel/Mobileye~\cite{mobileye}.
They are all investing billions of dollars and significant research efforts in this direction.
Even Xilinx, which traditionally targeted reconfigurable computing and signal processing systems, recently released its high-performance Ultrascale+~\cite{ultrascale-zcu102} System-on-Module that couples a multi-core host, a graphics co-processor and reconfigurable logics accelerator.
Their next platform generation, named Versal~\cite{versal}, is even more promising.

\textbf{Embedded Deep Neural Networks}.
Building future AD systems that are safe and reliable rises a number of challenges, because an autonomous vehicle requires to process a huge amount of data coming from cameras, LiDARs, radars and even the internet, to answer to questions such as: \textit{Where am I? Where is everybody else? How do I get from A to B?}, but also: \textit{What is the health status of the driver? Do I detect a possible danger for this vehicle, such as a crash with other vehicles or pedestrians?}.
The current way of answering these questions employs Deep Neural Networks (DNNs) at their state-of-the-art.
DNNs have a tremendous effectiveness for tasks as object detection and categorization, which are key modules of the perception and localization systems of tomorrow automated vehicles.
For instance, in traffic sign classification, the accuracy of an adequately trained DNNs can reach up to 99\%~\cite{traffic_sign}, even better than humans!

%
\begin{figure}[th!]
	\includegraphics[width=\columnwidth]{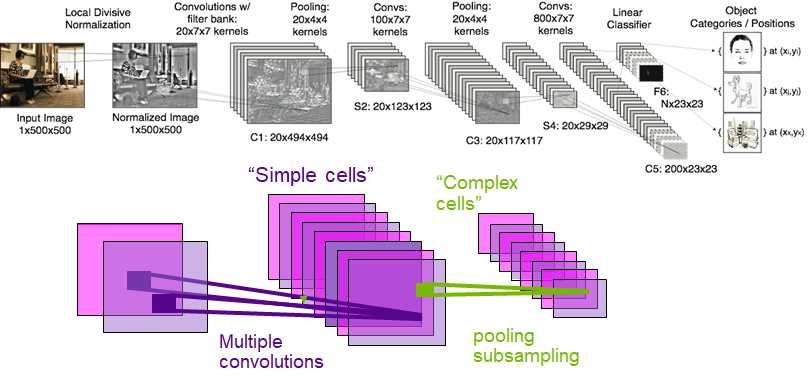}
	\centering
	\caption{Scheme of a generic CNN}
	\label{fig:cnn-scheme}
\end{figure}
%
To sustain the inherent complexity, hence, computational demand, of DNNs on size and power-constrained embedded platforms, typically system designers employ highly-optimized Convolutional Neural Networks (see Figure~\ref{fig:cnn-scheme}), where convolutional layers extract significant~ features from images.
Nicely, the main operations that CNNs carry are sums and multiplications, more specifically defined as \textit{Multiply-Accumulate Operations (MACs)}, which are highly scalable with the number of neurons, and data-parallelizable, hence perfectly suitable for deployment on many-core accelerators.

\textbf{What this paper is about}.
The goal of our research is to explore and assess the performance, accuracy and power profile of state-of-the-art CNN packages (es. YOLO~\cite{YOLO9000} and AlexNet~\cite{AlexNet}) for image detection and classification, when deployed on embedded computing platforms representative of next-generation automotive domain controllers.

We focus on the task of \textbf{inference}, and explore its limits and tradeoff when deployed on next generation of automotive systems, i.e., on tiny, embedded computers, under real-time constraints.
We do not focus on the (offline) training problem, and assume the network is already trained.
Nonetheless, we show how different datasets might affect performance of the same network on the same machine.

We choose three versions of the NVIDIA Tegra~\cite{xavier, jetson_tx2, nano}, as representative of GPGPU accelerators, and two variants of the Xilinx Ultrascale+~\cite{ultrascale-zcu102, ultra96}, for reconfigurable FPGA-based platforms.
For every CNN package, and target platform, we run an extensive set of experiments to assess the performance profile, not only in terms of computing power (where we choose frames-per-second and end-to-end latency on the single frame as representative metrics), but also the power dissipation at voltage and temperature operational ranges, in terms of consumed Watts.
Mostly all of these systems are battery-powered, and energy dissipation causes overheating, implying lower SoC lifetime and higher cooling costs.
Engineers struggle to find the optimal tradeoff between performance and dissipated power, and this is the reason why they typically measure the performance of these platforms in terms of \textit{TFLOPs-per-watt}.
To capture the power efficiency of the system with a signle metric for all the experiments, we introduce the FPS/Pow (of the sole SoC/SoM) metric.
For measuring network accuracy, we adopt the mean-Average Precision (mAP)~\footnote{We adopted the following tools to compute the mAP. For COCO: https://github.com/cocodataset/cocoapi, for all other tests: https://github.com/Cartucho/mAP}, which is a widely adopted and acknowledge way of assessing network precision.

This work follows up, and extends, our preliminary exploratory paper~\cite{brilli_HPCS18}, making the best of the results and observations achieved, which ultimately paved the way to this final and complete work.
More in details, in our previous work we generically focused on detection and classification CNN, with no in-depth analysis, and with a limited set of (now, obsolete) platforms and networks.
Here, we complete our analysis, and go deeper in exploring especially detection networks, building a comprehensive view of the problem.
More in details:\\
\textit{1)}. We re-computed our set of experiments on the next generation of computing platforms, including NVIDIA Xavier AGX~\cite{xavier} and the small Jetson NANO~\cite{nano} and Xilinx Ultra96~\cite{ultra96}. Few previous results (the ones on the ``old'' NVIDIA TX2) are still reported in Section~\ref{sec:arch-exploration} for the sake of comparison.\\
\textit{2)}. We extended our analysis to more advanced networks, mainly focusing on the object detection problem. We employ the newest YOLO v3 network~\footnote{At the time we write, also YOLO v4 and v5 are available, but still not stable nor fully supported on all of our reference platforms (for instance, on Xilinx's), hence we cannot perform a fair comparison}.\\
\textit{3)}.  We carried on a detailed, in-depth analysis of how specific features of the network or computing platform affect our metrics, in Section~\ref{sec:network-specific}. We focus on i) different datasets, ii) network quantization, iii) reduced ``tiny'' version of the same network, iv) image-level parallelism via batching (on NVIDIA architectures) or multiple threading (on Xilinx's), v) available power profiles via frequency and power scaling (available only on NVIDIA's GPGPUs -- Section~\ref{subsec:power-profiles}).

Section~\ref{sec:related_works} some previous works that are related to ours, and gives a brief motivation for the chosen network packages.
Section~\ref{sec:architecture} depicts the architectures considered in this work, and Section~\ref{sec:neural_networks} shows in details the properties and structure of the considered neural networks, and puts previous validation results in a line.
Sections~\ref{sec:arch-exploration} and~\ref{sec:network-specific} provide our detailed analysis of such CNN packages on the chosen reference platforms.
Finally, Section~\ref{sec:conclusions} draws some conclusions, and highlights future works.

%% file: related_works-full.tex
\section{Related works, and novelty}
\label{sec:related_works}

While neural networks are known since decades, doing inference on embedded platform is quite a recent application area, enabled by the IoT era with powerful mobile devices, and advances in automotive and avionics sectors.
Every time a new NN package comes to light, authors and interested researchers in the domain carry on an accurate evaluation to assess its advantages and disadvantages, on platforms that are representative of the specific application.


Recently, NVIDIA proposes two articles \cite{inference_bench_xavier}~\cite{xavier_tests} in which are reported benchmarks of the new Xavier chip compared to the previous version (TX2). The first one contains an exhaustive comparison of the main Convolutional Neural Networks on Jetson AGX Xavier.
The tests are made varying the power profile of the Xavier SoC: \textit{10W}, \textit{15W}, \textit{30W} and \textit{EDP} modes (trough the \textit{nvpmodel} tool) and varying the size of the batch loaded on the GPU.
This comparison is well presented but only compares the Xavier chip and does not consider the whole frequency range allowed by the JetPack.
The second article proposed by NVIDIA compares some CNNs on Xavier and TX2, but does not consider others architectures like FPGAs or Many-Cores.

J. Johnson proposes in \cite{bench_git} an analysis that involves some popular CNN models. In particular the results gathered by the author are carried out on several desktop GPUs, like: \textit{Pascal} and \textit{Maxwell Titan X} and \textit{GTX 1080} and \textit{1080 Ti}.
This paper differs from what J. Johnson published, because it is focused on desktop GPUs and does not take into account the power consumption.

Eriko Nurvitadhi et al. proposes in \cite{intel01} a customizable DNN template for FPGA and evaluates some emerging DNN algorithms on Intel Programmable Logics (Arria 10 and Stratix 10) against a desktop powerful GPU: the Pascal Titan X.
The results achieved by the authors show that the Stratix 10 is 10\%, 50\%, and 5.4x better in performance (TOP/sec) than Titan X Pascal GPU on GEMM operations for pruned, Int6, and binarized DNNs, respectively.

Nakahara et. al~\cite{BinarizedYOLOv2} developed an FPGA version of YOLO where inputs and weights are binarized \cite{Binarized}, this network is implemented with SDSoC, an high level synthesis tool provided by Xilinx.
They also carry an extensive benchmarking of their network running on Zynq Ultrascale+ and Tegra X2, the comparison shows that the FPGA implementation can reach about 40 frames per second (FPS) with a power dissipation of 4.5 Watt, while the GPU implementation that runs on Tegra X2 achieves only 2 FPS with a consumption of 7 Watt.
These latter numbers can certainly be improved.

The Xilinx engineer M. Blott proposed a reduced and quantized version of Tiny-YOLO called Tincy-YOLO~\cite{TincyYOLO}, which operates at 16 FPS, consuming only 6 Watt for the FPGA accelerator.
The benchmark proposed in the article regards a comparison of the accuracy achieved by Tincy-YOLO respect to other versions of YOLO, but does not include tests with others architectures like GPUs.
We aim at completing these results.

Our aim is to assess i) \textbf{both} performance and power efficiency, ii) of \textbf{multiple} state-of-the-art neural network packages iii) on \textbf{multiple} platforms, which are representative of the future embedded systems, not only in the automotive domain (as for positioning of this paper), but also on the embedded domain \textit{tout court}.
We believe this work can be a reference for system engineers, to assess and identify the most suitable packages for given hardware system, and the other way around.

Kim et al.~\cite{kim_18} perform a comparison of embedded deep learning methods for person detection on embedded platforms such as the Jetson TX2 and Movidius. They investigate the performance of 13 different object detector deep models including variants of YOLO, SSD, RCNN, R-FCN and SqueezeDet.
They train and test the NN on their proprietary dataset. In order to deploy the deep models in embedded platforms, they optimize Caffe or Tensorflow implementation using Movidius SDK or TensorRT. This enables the CNN model to utilize the target height/width effectively. The execution times of the methods are computed on the embedded boards. On the other hand, to measure and compare the average precision (AP) and
IoU of the deep models, they used a workstation powered by 16 GB of internal memory and Nvidia GTX 1080ti graphics accelerator. Experiments results shows that Tiny YOLO-416
and SSD (VGG-300) are among the fastest models and Faster RCNN (Inception ResNet-v2) and RFCN (ResNet-101) are the most accurate ones. YOLO v3-416 and SSD (VGG-500) are the best tradeoff between Average precision and throughput.

Mittal presents in~\cite{mittal_18} a survey of works that evaluate and optimize neural network applications on Jetson platform (TK1, TX1, TX2). The survey considers several computer vision applications and offers a good overview of the existing literature. However, it does not take into account the Xavier platform.

Yu et a.~\cite{yu_18} compare Real-Time object detection algorithms on embedded platforms. They compare power efficiency,
latency and accuracy of Faster RCNN, YOLO and SSD on NVDIA TK1, XILINX Zynq 7045 and XILINX KU115.
However the comparison is not very clear, and different settings are compared, making the comparison unfair.

Chen and Ran~\cite{chen_19} present a review of deep learning with edge computing. They report background, measurements and frameworks, consider different applications of deep learning, from computer vision to natural language processing, and take into account several methods for fast inference. Regarding fast inference on device computation, three major efforts have been identified:
- Model Design: such methods focus on designing models with a reduced number of parameters in the DNN model, thus reducing memory and execution latency, while aiming to preserve high accuracy.
Some examples include MobileNets \cite{howard2017mobilenets}, Single-Shot Detector (SSD)\cite{liu2016ssd}, YOLO\cite{YOLO9000}\cite{yolov2}, and SqueezeNet \cite{SqueezeNet}, with the state of the art that is evolving rapidly.
- Model Compression: such methods seek to compress the existing DNN models with minimal accuracy loss compared with the original model. There are several popular model compression methods: parameter quantization, parameter pruning, and knowledge distillation.
- Hardware: To speed up inference of deep learning, hardware manufacturers are leveraging existing hardware such as CPUs and GPUs, as well as producing custom application-specific integrated circuits (ASICs) for deep learning. As an example, consider NVIDIA GPUs, FPGAs or Google TPUs.

Hossain and Lee~\cite{hossain_19} design a deep learning Real-Time object detection and tracking method for drones with Embedded Device. In this work they compare several SOTA NN for object detections on some GPU based embedded boards. In particular, they consider YOLOv2, YOLOv3, tiny YOLO, SDD and Faster RCNN on TX1, TX2 and AGX Xavier. Moreover, they take into account YOLOv2 and SSDMobileNet on Raspberry Pi, Latte Panda, Odroid, with or without Movidious. The comparison is only performed con the execution times of the neural networks.

Rungsuptaweekoon et al.~\cite{rungsuptaweekoon_17} evaluate the power efficiency of image recognition with YOLO on NVIDIA Jetson TX1, Jetson TX2, and Tesla P40. For this evaluation, they deployed the Low-Power Image Recognition Challenge (LPIRC) system and integrated YOLO, a power meter, and target hardware into the system. The authors compare mAP, accumulated energy consumption, mAP/Energy and frame rate factors. The experimental results show that Jetson TX2 with Max-N mode has the highest throughput; Jetson TX2 with Max-Q mode has the highest power efficiency.

Ding et al.~\cite{ding_19} propose REQ-YOLO, a resource aware, systematic weight quantization framework for object detection, considering both algorithm and hardware resource aspects in object detection. They consider Yolo v2 tiny and several boards: NVIDIA Titan X, GTX 1070, Jetson TX2 and Xilinx Virtex-7 485t, Zynq 7020, ADM-7V3 FPGA. They compare the performance on the different platforms in terms of FPS and power (W). They show that, thanks to their framework, they can achieve 314.2 FPS using only 21 W on the ADM-7V3 FPGA.

Lin et al.~\cite{lin_19} benchmark several deep learning frameworks (MXNet\cite{mxnet}, TensorFlow\cite{tensorflow}, CNTK\cite{cntk}, Neon\cite{neon}, and PyTorch\cite{pytorch}) and investigate the FPGA deployment for performing traffic sign classification and detection. They evaluate both training and inference performance. For the latter one, they consider inference speed, accuracy, and power efficiency, by varying different parameters such as floating point precision, batch sizes, etc. They show that TensorFlow is always among the frameworks with the highest inference accuracy. For object detection inference, they compare six SSD models with different base networks, namely, ResNet-18, ResNet-50 \cite{he2016deep}, MobileNet-v1 \cite{howard2017mobilenets}, SqueezeNet-v1.1 \cite{SqueezeNet}, VGG \cite{simonyan2014very}, and MobileNet-v2 \cite{sandler2018mobilenetv2} (with SSDLite), on an NVIDIA GTX 1050 Ti GPU and an Arria 10 FPGA development board. Compared to the reference results on the GPU, they notice that in most of the cases inference speed on the GPU is higher than the FPGA, as well as accuracy. However, FPGA always achieves higher power efficiency than the GPU. Among the FPGA test cases, FP11 is always more power-efficient than FP16, while the data type does not have a clear impact on the power efficiency, but using FP11 bitstreams results in some decrease in accuracy. Finally, if the bitstream being used is fixed, the data type makes only a very tiny difference on both inference speed and accuracy.

%% file: architectures.tex
\section{Automotive Architectures}
\label{sec:architecture}
\begin{figure}
	\includegraphics[width=.8\columnwidth]{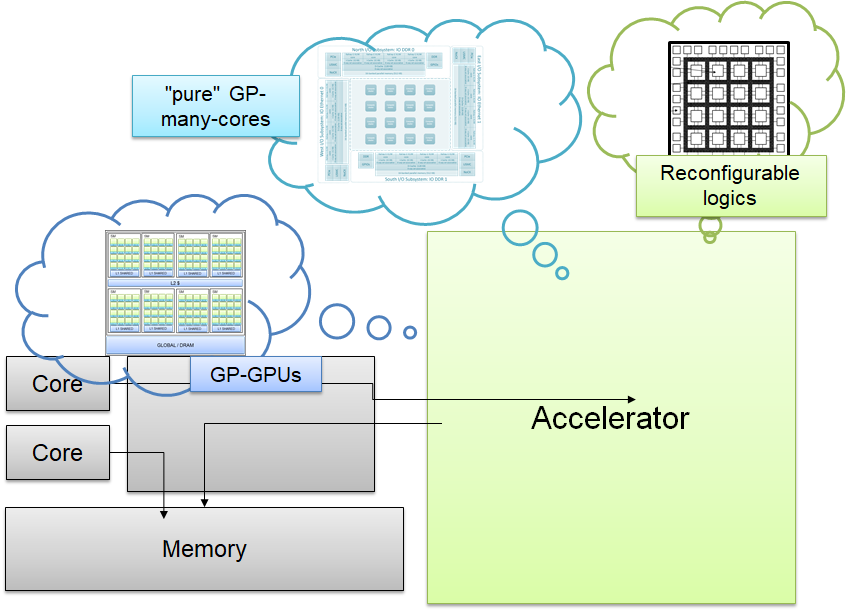}
	\centering
	\caption{Modern embedded platform for NN acceleration}
	\label{fig:emb-platforms}
\end{figure}

In this work, we target a quite common architectural template for embedded systems, the \textit{host-Accelerator paradigm}
, where a multi-core host is coupled with a power-efficient many-core accelerator (See Figure~\ref{fig:emb-platforms}).
Application is partitioned so that the host subsystem manages data transfer, in our case, the frames from the camera and the accelerator/system memory, and the many-core accelerator runs the highly-parallel workload, that is, in our case, the convolutional neurons, in an energy-efficient manner.
We target several different architectural ``flavors'', namely three based on NVIDIA GPGPUs accelerators, and two based on reconfigurable FPGA logics.
They currently represent the most advanced technologies available on the market in each domain.

\textbf{GP-GPU: the NVIDIA Tegra family and AGX Xavier}
The \textit{Tegra TX2} board embeds an esa-core host with Big.SUPER configuration and a GPU with 2 NVIDIA Streaming Multiprocessors (SM) of the Pascal family, summing up to 256 CUDA cores.
The host and accelerators communicate using shared memory banks, employing what NVIDIA calls the \textit{Unified Virtual Memory} space.
The platform is claimed to achieve 1 TFLOP of computing power, within approximately 20 Watts of power dissipation.
Being explicitly designed for the automotive market, Tegra X2 is qualified according to Functional Safety and Road Vehicles Standard (ISO 26262's ASIL-B level~\cite{iso26262}), and has been marketed in a (family of product) named Drive CX/PX~\cite{drive_px, drive_px_gtc}.

\textit{Xavier} is the new chip developed by NVIDIA with even more computational power.
It includes an integrated Volta GPU composed by 512 CUDA-Cores with 64 Tensor Cores, a dual Deep Learning Accelerators (NVDLAs), an octal-core NVIDIA Carmel ARMv8.2 CPU, 16GB 256-bit LPDDR4x with 137GB/s of memory bandwidth, and 650Gbps of high-speed I/O including PCIe Gen 4 and 16 camera lanes of MIPI CSI-2.
According to NVIDIA this chip is able to reach a peak performance of 32 TOPS under 30W.

We also consider family's low-end platform, the Jetson \textit{NANO}~\cite{nano}, a very small compute module that embeds a quad-cores ARM A57 processor, a GPGPU with only one SM composed of 128 CUDA cores and a 4GB of LPDDR4 main memory.
Unfortunately, due to its constrained resources, it does not support bigger networks, such as full YOLO.

\textbf{Reconfigurable logics: the Xilinx Ultrascale+}
The second platform family we consider is Xilinx Zynq UltraScale+ (XU+), a next-generation reconfigurable heterogeneous platform that couples power-efficient host complex (ARM A53 cores with Real-Time cores, the ARM Cortex R5), and reconfigurable logics.
The SoC also features a Mali GPU, that anyway can be used only for graphics.
In our setup we used two different versions of the XU+ family: the XCZU9EG and the XCZU3EG.
%
For the XCZU9EG SoC we used the Xilinx Zynq \textit{ZCU102}~\cite{ultrascale-zcu102} a very big carrier board with a really rich I/O connectivity, that can become quite energy-consuming, for this reason we expect its power consumption to be higher than the other ones.
For the XCZU3EG, we adopted the Avnet \textit{Ultra96} development platform \cite{ultra96} a carrier board with a very small form factor, for the latter we expect a power consumption close to Jetson Nano.
The XCZU3EG have about the $25\%$ of the total reconfigurable resources of the XCZU9EG.

\textbf{Expected Results}.
We expect to see better performance on GPGPUs in general, mainly due to the fact that the frameworks used are widely tested and therefore highly optimized.
More in details, we expect an increase of about $3\times$ factor across different platform generations, since NVIDIA declares a computational power of about $1$ TFLOP for TX2 and about $2.8$ TFLOPS for Xavier.
With regard to the performance on FPGA-based architectures, we expect lower throughput in terms of FPS due to the lower diffusion of the aforementioned chips and the lower life cycle of the frameworks used.
Finally, regarding the power consumption, we expect lower consumption on FPGA, and an excellent energy efficiency on XU+ and NVIDIA Xavier.

%% file: neural_networks.tex
\section{Convolutional neural networks}
\label{sec:neural_networks}
%
Convolutional Neural Networks or \textit{CNNs} are typologies of neural networks particularly suitable for computer vision, image recognition and in general for processing 2D-input data. The basic structure of a CNN consists of a stack of layers in which an input image flows from the input layer toward the output layer.
The input image is called \textit{input feature map (fmap)} and it is mapped directly into the input neurons; this is beneficial in image recognition, because the spatial information inside images is preserved.
The input feature map is represented as an object in three dimensions called \textit{tensor}, whose dimensions are height $H_{in}$, width $W_{in}$ and depth $D_{in}$.
Each layer of which the CNN is composed, produces a stack of $D_{out}-$images of $W_{out} \times H_{out}-$pixels, this output tensor is called \textit{output feature map}.
The most important and one of the most compute-intensive operation performed by CNNs is the discrete 2D-convolution, it is performed by convolutional layers and it is defined as:

$$
	o_{i, j}^k = \sum_{c = 0}^{D_{in}} \sum_{h = 0}^{K_H} \sum_{w = 0}^{K_W}(w_{h, w, c}^kx_{i+h, j+w, c}) + b_k
$$

Due to its mathematical structure, the convolutional layers, are suitable to be implemented on parallel hardware like Many-Cores platforms.

%
Image Classification is probably the most well-known problem in computer vision and it consists of classifying an image into one or more different categories.
The problem of Object Detection is more complex, as it consists of classifying several objects inside an input image with some spatial information called \textit{bounding boxes}, and for this reason, this operation requires more computing power and energy to be carried out.

\textbf{Detection: You-Only-Look-Once (YOLO)}.
The model of the CNNs for objects detection analyzed in this paper is the YOLO~\cite{YOLO9000}, the state of the art for objects detection via Convolutional Neural Networks. Almost the whole detection systems apply its model to an image at multiple locations and scales and then returns the computed bounding boxes that achieves the maximum scores. YOLO uses a totally different approach, it applies a single neural network to the full input image, the network divides the image into smaller regions and predicts bounding boxes and classes for each region, the network also compute weights for all regions and finally it returns only the bounding boxes which exceeds a fixed
threshold. The YOLO full model requires a lot of memory and computations, then the authors proposed some different and reduced versions of the original model called YOLO-Small and Tiny-YOLO.

\textbf{Classification: AlexNet}.
AlexNet is a large scale deep convolutional neural network proposed by A. Krizhevsky et al. in [1], AlexNet is designed for image classification and trained on ImageNet ILSVRC.
This CNN is composed by eight trained layers: five convolutional layers and three fully connected layers, the network starts performing an image reduction to match the size of the input neurons that is $227\times227\times3$.
AlexNet introduces some new layers, for example a new normalization layer called \textit{Local Response Normalization (LRN)}
.

%% file: experiments-hpcs.tex
\section{Architectural exploration}
\label{sec:arch-exploration}
%
%

We recall that we are in the automotive domain, where systems not only deliver high peek performance, but also do it in a power efficient manner.
For this reason, we measure performance of considered CNNs under two major metrics, that are, the \textit{number of frames} processed in one second and the \textit{total power} dissipated by the SoC in the inference phase.
We also introduce a last metric that is composed by FPS and power, hence captures the \textit{power-efficiency} of the system.
Unless specified, all throughput measures are in Frames-per-Second, latency measures are in milliseconds, and power consumption is in Watt.

%
Since NVIDIA boards provide frequency scaling mechanism both for the host/CPU and accelerator/GPGPU, for Tegra X2 and Xavier we show 3D graphs where the x-axis reports the GPU frequency ranging from about 100MHz up to 1,3GHz.
In the y-axis is reported the whole host frequencies, that ranging from about 300MHz up to about 2GHz.
Finally, the metric of interest is reported in z-axis.

For the Host part, here we have a single core for feeding video frames to the network and one core for reading the values of the sensor; the others cores are off.
Multi-threading and batching are explored later, in Section~\ref{subsec:batching-threading}.
We don't want our experiments to be disturbed by software interference from other application, and for this reason we set the \textit{realtime} priority of the task that runs the network to the maximum, using (the native) Ubuntu Linux API \texttt{chrt}, so we avoid unwanted context switches and through \texttt{taskset -c} we set a cpu affinity for the task that handles the network and the one that read the sensor.\\

\textbf{Classification: AlexNet}.
In the first comparison we show the AlexNet model, that as mentioned above is a CNN for image classification.
\begin{figure}[t!]
	\includegraphics[width=0.8\columnwidth]{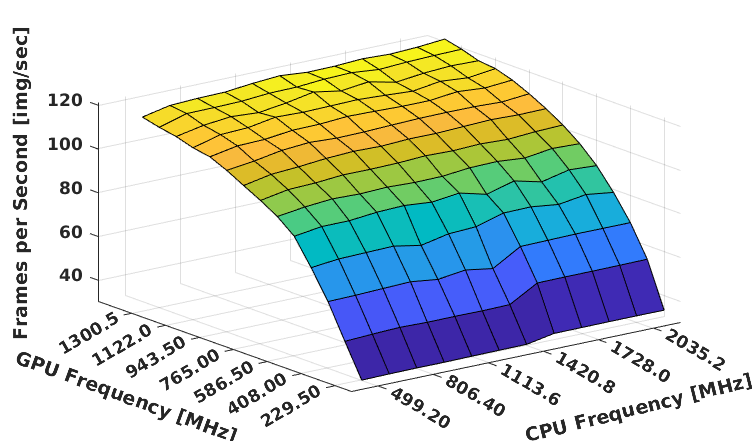}
	\centering
	\caption{AlexNet on TX2, $FPS$ @ 19V}
	\label{AlexNetTX2FPS}
\end{figure}
\begin{figure}[t!]
	\includegraphics[width=0.8\columnwidth]{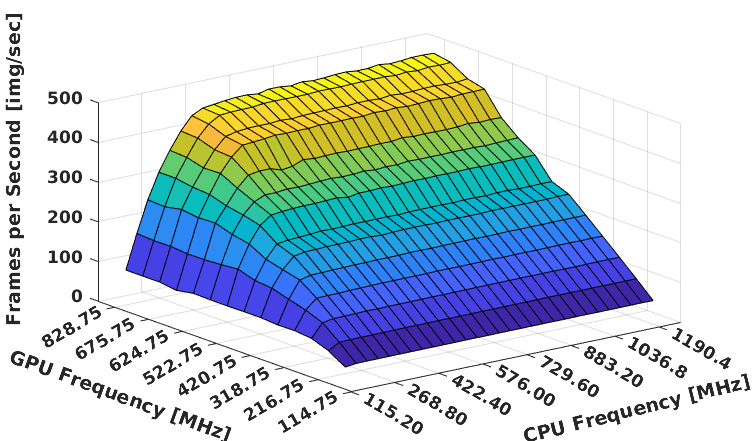}
	\centering
	\caption{AlexNet on Xavier, $FPS$ @ 19V}
	\label{AlexNetXavierFPS}
\end{figure}
%
%
\begin{table}[t!]
	\centering
    \begin{adjustbox}{width=\columnwidth}
    	\begin{tabular}{c|c|c|c|c|c}
            \textbf{PS Freq.} & \textbf{PL Freq.} & \textbf{T.Put} & \textbf{Pboard} & \textbf{Pinf}   & \textbf{Esys}      \\
            \textbf{[MHz]}    & \textbf{[MHz]}    & \textbf{[FPS]} & \textbf{[Watt]} & \textbf{[Watt]} & \textbf{[FPS/Watt]}\\
            \midrule
            2400 & 200  & 61.554 & 23.1165 & 0.5097 & 120.7652\\
            1110 & 99   & 0.2462 & 20.2231 & 0.4603 & 0.534900\\
            \bottomrule
    	\end{tabular}
    \end{adjustbox}
    \label{AlexNetUltraScale}
    \caption{AlexNet on Zynq UltraScale+}
    \footnotesize
\end{table}
Figures \ref{AlexNetTX2FPS}, \ref{AlexNetXavierFPS}, and \ref{AlexNetUltraScale} show the frames-per-second achieved, respectively on TX2, Xavier, and UltraScale+.
For inference on this network, GPUs outperforms the other two accelerators by an order of magnitude, in particular the TensorRT implementation, surprisingly was able to reach up to about $120$ FPS on TX2 and about $300$ FPS on Xavier, while the other are more or less comparable: about $60$ FPS on UltraScale+.

It is interesting to note how, in case of TX2 platform (Figure \ref{AlexNetTX2FPS}), performance does not seem to be affected much by host frequency scaling, but only by GPGPU scaling, and this means that the CPU can keep the pace of the accelerator also at lower frequencies.
This is a positive effect on power efficiency, as shown in~\cite{brilli_HPCS18}\footnote{For reasons of space, we do not report here energy charts, as they are already published}.
Also, the GPGPU keeps good performance scaling with frequency, meaning that we can potentially run additional workload (in case, exploiting the missing ARM cores as controller) without performance loss.
Similarly, the Xavier board (Figure \ref{AlexNetXavierFPS}), on the other hand, shows a tremendous peek performance of $300$ FPS, which can potentially be improved increasing GPGPU frequency.
Also, here there is poor we can do increasing host core frequency.

\begin{figure}[t!]
	\includegraphics[width=0.8\columnwidth]{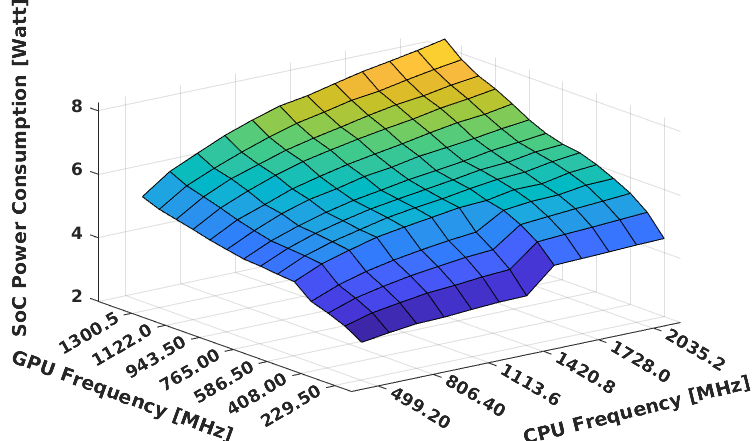}
	\centering
	\caption{AlexNet on TX2, $P_{SoC}$ @ 19V}
	\label{AlexNetTX2PowerSoC}
\end{figure}

\begin{figure}[t!]
	\includegraphics[width=0.8\columnwidth]{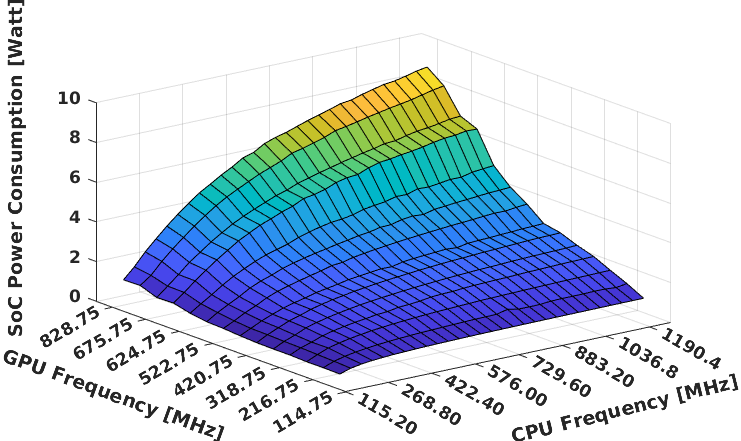}
	\centering
	\caption{AlexNet on Xavier, $P_{SoC}$ @ 19V}
	\label{AlexNetXavierPowerSoC}
\end{figure}

Regarding the power consumption, the $P_{SoC}$ parameter, Figures \ref{AlexNetTX2PowerSoC} and \ref{AlexNetXavierPowerSoC} show comparable performance on GPGPUs (about $1.5$ and $4.5$ Watts on average for TX2 and Xavier).
In figure \ref{AlexNetUltraScale} we report the power consumption$P_{inf}$ of the FPGA-based accelerator, that was able to run AlexNet with about $500$ milliWatts.

\textbf{Detection: The YOLO-family}.
In this section, we show results for the detection network YOLO in its reduced version Tiny-YOLO, that compared to the full model, has less memory footprint and less complexity.
\begin{figure}[t!]
	\includegraphics[width=0.8\columnwidth]{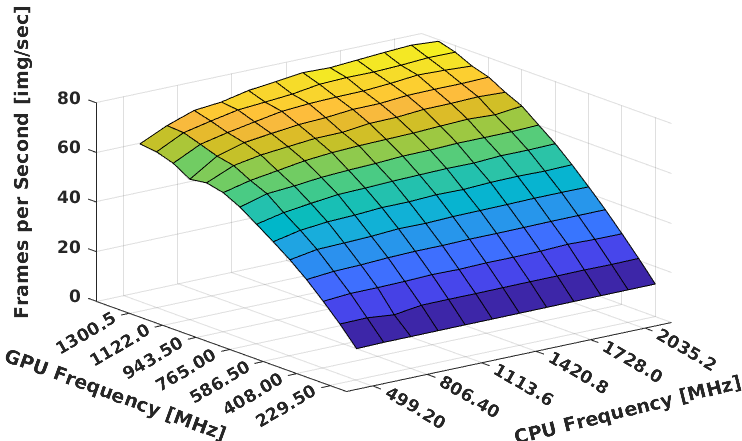}
	\centering
	\caption{Tiny-YOLO on TX2, $FPS$ @ 19V}
	\label{TinyYOLOTX2FPS}
\end{figure}
\begin{figure}[t!]
	\includegraphics[width=0.8\columnwidth]{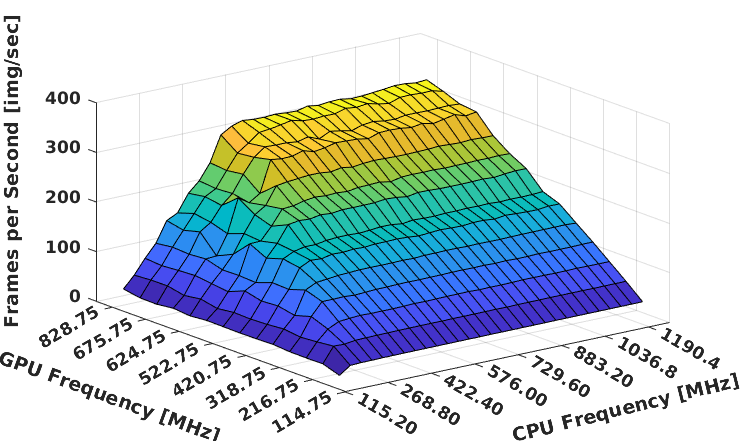}
	\centering
	\caption{Tiny-YOLO on Xavier, $FPS$ @ 19V}
	\label{TinyYOLOXavierFPS}
\end{figure}
%
%
\begin{table}[t!]
	\centering
    \begin{adjustbox}{width=\columnwidth}
    	\begin{tabular}{c|c|c|c|c|c|c}
            \multirow{2}{*}{\textbf{Model}} & \textbf{PS Freq.} & \textbf{PL Freq.} & \textbf{T.Put} & \textbf{Pboard} & \textbf{Pinf}   & \textbf{Esys}      \\
                                   & \textbf{[MHz]}    & \textbf{[MHz]}    & \textbf{[FPS]} & \textbf{[Watt]} & \textbf{[Watt]} & \textbf{[FPS/Watt]}\\
            \midrule
            Tiny-YOLO   & 2400 & 200 & 22.6807 & 24.1798 & 1.1465 & 19.7826\\
            Small-YOLO  & 2400 & 200 & 7.93110 & 23.7137 & 0.6804 & 11.6565\\
            YOLO        & 2400 & 200 & 6.67288 & 23.5545 & 0.5217 & 12.7906\\
            \bottomrule
    	\end{tabular}
    \end{adjustbox}
    \label{TinyYoloUltrascale}
    \caption{YOLOs on Zynq UltraScale+}
    \footnotesize
\end{table}
Figures \ref{TinyYOLOTX2FPS}, \ref{TinyYOLOXavierFPS}, and \ref{TinyYoloUltrascale}, show a big difference in terms of frame-per-second on GPUs compared to the others: in particular, this model on TX2 and Xavier can reach about $70$ and $300$ FPS respectively, while on UltraScale+ reach about $22$ FPS.
Regarding the power consumption. we have experimented the same trend as AlexNet: a power-hungry behavior on GPUs (about $7$ and $10$ Watts on average), to sustain the tremendous peek performance.
UltraScale+ still consumes about an order of magnitude less power respect to the others architectures.
\begin{figure}[t!]
	\includegraphics[width=0.8\columnwidth]{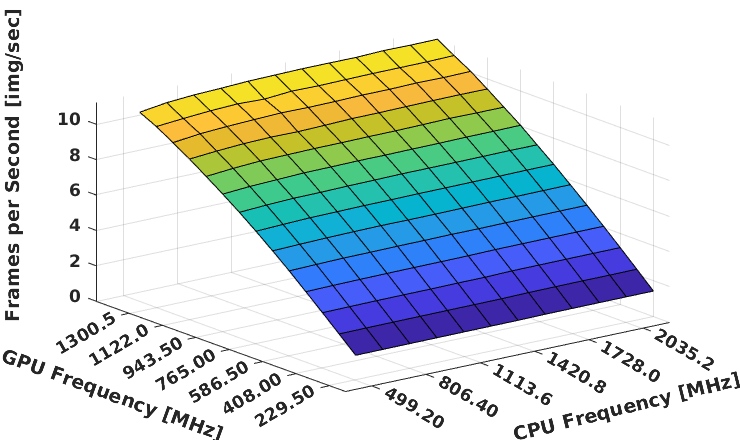}
	\centering
	\caption{YOLO on TX2, $FPS$ @ 19V}
	\label{YOLOTX2FPS}
\end{figure}
\begin{figure}[t!]
	\includegraphics[width=0.8\columnwidth]{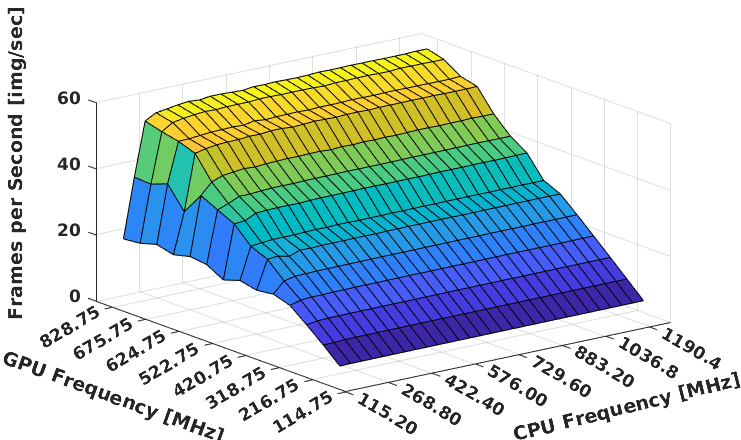}
	\centering
	\caption{YOLO on Xavier, $FPS$ @ 19V}
	\label{YOLOXavierFPS}
\end{figure}
In this case we have good performance on Xavier, about $55$ FPS (Figure \ref{YOLOXavierFPS}) and comparable performance on TX2 and XU+: about $11$ and $7$ FPS respectively (Figures \ref{YOLOTX2FPS} and \ref{TinyYoloUltrascale}).

\textbf{Reasoning on results}.
In general, in terms of FPS, we can see that beyond particular frequencies of the CPU, the number of frames processed per second does not increase much, this makes us understand that in terms of power-efficiency it is not convenient to maximize the frequency of the processor, but there are fixed speeds in which the power-efficiency is maximized.
For example, with reference to Figures \ref{YOLOTX2FPS} and \ref{YOLOXavierFPS} (where the frames processed by the YOLO network on TX2 and Xavier are shown) we can see that, on TX2, scaling the host core from about $350$MHz to $2035$MHz we have an increase of really few FPS, about $2-3$.
On Xavier (shown in Figure \ref{YOLOXavierFPS}) we have the same behavior, supposing to not consider the $115$, $192$ and $268$MHz frequencies that are excessively low to be able to feed the accelerator constantly with video frames.
This behavior can be seen mainly on computationally-intensive neural networks, such as \textit{Tiny-YOLO} and even more \textit{YOLO}.

\begin{figure}[!ht]
	\includegraphics[width=0.8\columnwidth]{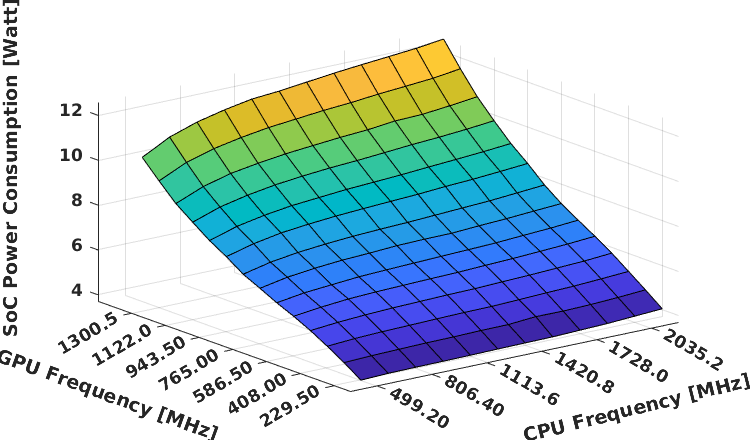}
	\centering
	\caption{Tiny-YOLO on TX2, $P_{SoC}$ @ 19V}
	\label{TinyYOLOTX2PowerSoC}
\end{figure}

\begin{figure}[t!]
	\includegraphics[width=0.8\columnwidth]{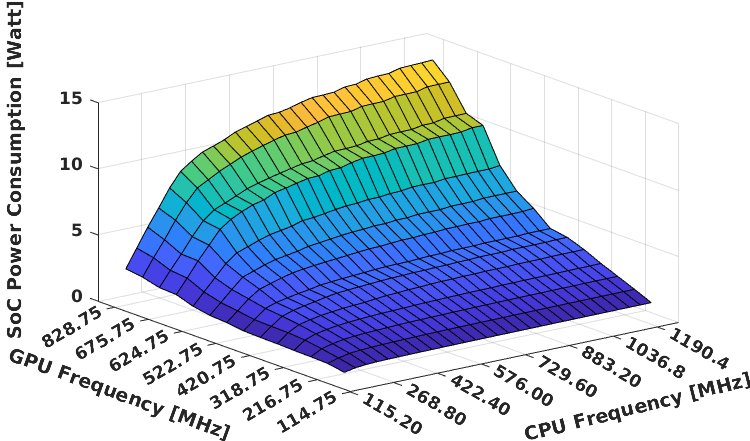}
	\centering
	\caption{Tiny-YOLO on Xavier, $P_{SoC}$ @ 19V}
	\label{TinyYOLOXavierPowerSoC}
\end{figure}

\begin{figure}[t!]
	\includegraphics[width=0.8\columnwidth]{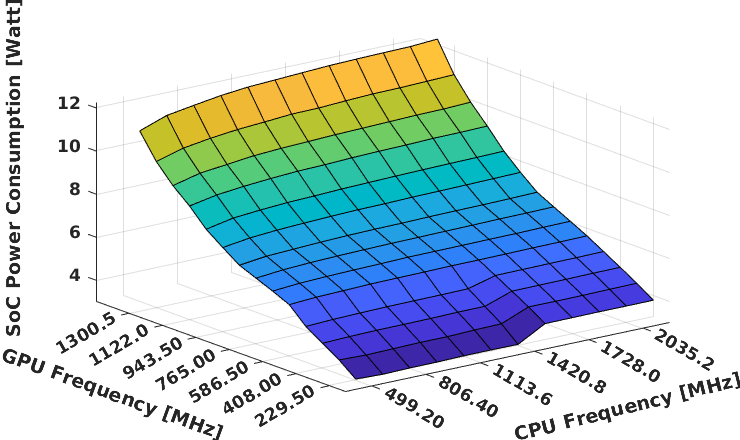}
	\centering
	\caption{YOLO on TX2, $P_{SoC}$ @ 19V}
	\label{YOLOTX2PowerSoC}
\end{figure}

\begin{figure}[t!]
	\includegraphics[width=0.8\columnwidth]{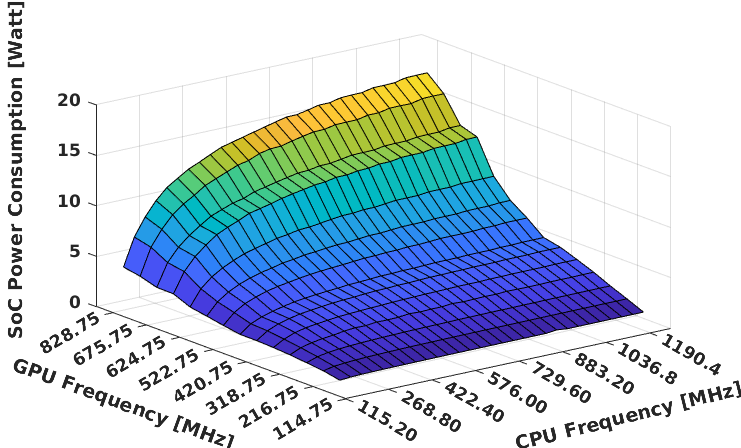}
	\centering
	\caption{YOLO on Xavier, $P_{SoC}$ @ 19V}
	\label{YOLOXavierPowerSoC}
\end{figure}

The charts in Figure~\ref{TinyYOLOTX2PowerSoC} to~\ref{YOLOXavierPowerSoC} show the SoC power consumption of inference phase of YOLO variants on the considered platforms.

\textbf{Observation}:\textit{For the reasons mentioned above, it is not always convenient to maximize the frequencies and in this way the power-efficiency surface takes an inverted 3D parabolic shape.}

Discussing about the accelerator frequency, we will discuss only the GPGPU case, because only Tegras expose a reasonably high number of frequency slots to let results be representative.
In this case, with reference to the power graphs, for example \ref{YOLOTX2PowerSoC} and \ref{YOLOXavierPowerSoC}, (where we can see the power dissipated by the \textit{YOLO} network) we notice a steep growth in power consumption at high accelerator frequencies.
In terms of FPS, on the other hand, we have noticed an opposite trend, and the framerate achieved increases slowly at high frequencies.
This behavior leads to a maximum point in the power-efficiency function, in which the accelerator frequency is not at its higher value.

\subsection{Power profiles (GPGPU only)}
\label{subsec:power-profiles}
Besides core frequency scaling, NVIDIA GPGPUs offer a very powerful mechanism, that are, \textbf{power models}.
This platform-specific feature enables different CPU-GPU frequency scaling in a programmer-friendly manner, without having to manually tune their cores' frequencies.
They are shown in Tables~\ref{tab:powermodes-yolov2} and~\ref{tab:powermodes-yolov3}.
For all platforms, '0' is the most performing, hence power expensive, mode, while '1' is the other way around.
Other models numbers, where available, are in between this spectrum, in increasing performance.
We carried on an extensive set of experiments on two networks (namely, Yolo v2 and Yolo v3) on the three reference boards.
Due to reasons of space, we report in Tables only results for the COCO dataset (and not YoloV2-tiny), but the trend is similar for all networks and datasets, and this is probably the most interesting result.

\textbf{Observation}: \textit{power modes don't seem to significantly affect network precision (see mAP column), and do not significantly impact on any network or dataset more than others}.
%
%
\begin{table*}[t!]
	\centering
    \begin{adjustbox}{width=\textwidth}
    	\begin{tabular}{l|cccccc|cccccc|cccccc}
            & \multicolumn{6}{c|}{\textbf{TX2}} & \multicolumn{6}{c|}{\textbf{Xavier}} & \multicolumn{6}{c}{\textbf{Nano}}\\
            \bottomrule
            
            \multirow{2}{*}{\textbf{Profile}} & \textbf{E2E} & \multirow{2}{*}{\textbf{T.Put}} & \textbf{Pow} & \textbf{Pow} & \multirow{2}{*}{\textbf{mAP}} & \multirow{2}{*}{\textbf{P.eff}} & \textbf{E2E} & \multirow{2}{*}{\textbf{T.Put}} & \textbf{Pow} & \textbf{Pow} & \multirow{2}{*}{\textbf{mAP}} & \multirow{2}{*}{\textbf{P.eff}} & \textbf{E2E} & \multirow{2}{*}{\textbf{T.Put}} & \textbf{Pow} & \textbf{Pow} & \multirow{2}{*}{\textbf{mAP}} & \multirow{2}{*}{\textbf{P.eff}}\\
            
            & \textbf{latency} &  & \textbf{(board)} & \textbf{(SoC)} &  &  & \textbf{latency} &  & \textbf{(board)} & \textbf{(SoC)} &  &  & \textbf{latency} &  & \textbf{(board)} & \textbf{(SoC)} &  & \\
            \midrule
            0 (perf) & 145.10 & 6.88 & 12.54 & 8.873 & 23.90\% & 0.775 & 62.07 & 16.11 & 37.354 & 26.98 & 23.90\% & 0.597 & 405.39 & 2.466 & 8.9303 & 6.7476 & 0.239 & 0.365\\
            1 (power) & 228.70& 4.37 & 6.346 & 3.952 & 23.90\% & 1.106 & 264.75 & 3.777 & 11.438 & 3.6746 & 23.90\% & 1.028 & 586.48 & 1.705 & 5.2426 & 3.1873 & 0.239 & 0.535\\
            2 & 176.70 & 5.65 & 6.194 & 5.605 & 23.90\% & 1.008 & 121.06 & 8.26 & 15.8422 & 7.8147 & 23.90\% & 1.057 & - & - & - & - & - & - \\
            3 & 164.10 & 6.09 & 7.125 & 6.574 & 23.90\% & 0.926 & 97.17 & 10.29 & 20.2578 & 11.6641 & 23.90\% & 0.882 & - & - & - & - & - & -\\
            4 & 159.80 & 6.25 & 7.201 & 6.631 & 23.90\% & 0.943 & 94.39 & 10.59 & 20.9475 & 12.3842 & 23.90\% & 0.855 & -& -& -& -& -& -\\
            5 & - & - & - & - & - & - & 91.21 & 10.96 & 21.1356 & 12.5894 & 23.90\% & 0.871 & - & - & - & - & - & -\\
            6 & -& -& -& -& -& - & 89.40 & 11.19 & 21.2648 & 12.7357 & 23.90\% & 0.879 & -& -& -& -& -& -\\
            \bottomrule
    	\end{tabular}
    \end{adjustbox}
    \label{tab:powermodes-yolov2}
    \caption{Powermodes on Yolo v2}
    \footnotesize
\end{table*}
%
%
\begin{table*}[t!]
	\centering
    \begin{adjustbox}{width=\textwidth}
    	\begin{tabular}{l|cccccc|cccccc|cccccc}
            & \multicolumn{6}{c|}{\textbf{TX2}} & \multicolumn{6}{c|}{\textbf{Xavier}} & \multicolumn{6}{c}{\textbf{Nano}}\\
            \bottomrule
            
            \multirow{2}{*}{\textbf{Profile}} & \textbf{E2E} & \multirow{2}{*}{\textbf{T.Put}} & \textbf{Pow} & \textbf{Pow} & \multirow{2}{*}{\textbf{mAP}} & \multirow{2}{*}{\textbf{P.eff}} & \textbf{E2E} & \multirow{2}{*}{\textbf{T.Put}} & \textbf{Pow} & \textbf{Pow} & \multirow{2}{*}{\textbf{mAP}} & \multirow{2}{*}{\textbf{P.eff}} & \textbf{E2E} & \multirow{2}{*}{\textbf{T.Put}} & \textbf{Pow} & \textbf{Pow} & \multirow{2}{*}{\textbf{mAP}} & \multirow{2}{*}{\textbf{P.eff}}\\
            
            & \textbf{latency} &  & \textbf{(board)} & \textbf{(SoC)} &  &  & \textbf{latency} &  & \textbf{(board)} & \textbf{(SoC)} &  &  & \textbf{latency} &  & \textbf{(board)} & \textbf{(SoC)} &  & \\
            \midrule
            0 (perf) & 291.48 & 3.43 & 15.105 & 11.438 & 31.80\% & 0.3 & 106.59 & 9.38 & 38.399 & 28.025 & 31.80\% & 0.335 & - & - & - & - & - & - \\
            1 (power) & 431.77 & 2.32 & 7.904 & 5.51 & 31.80\% & 0.4 & 514.25 & 1.944 & 11.362 & 3.5986 & 31.80\% & 0.540 & - & - & - & - & - & - \\
            2 & 332.70 & 3.00 & 11.438 & 10.849 & 31.80\% & 0.3 & 211.50 & 4.72 & 16.283 & 8.2555 & 31.80\% & 0.572 & - & - & - & - & - & - \\
            3 & 333.63 & 2.99 & 11.362 & 10.811 & 31.80\% & 0.3 & 163.60 & 6.11 & 21.47 & 12.8763 & 31.80\% & 0.475 & - & - & - & - & - & - \\
            4 & 334.56 & 2.98 & 11.248 & 10.678 & 31.80\% & 0.3 & 163.35 & 6.12 & 21.337 & 12.7737 & 31.80\% & 0.479 & - & - & - & - & - & - \\
            5 & - & - & - & - & - & - & 161.45 & 6.19 & 21.66 & 13.1138 & 31.80\% & 0.472 & - & - & - & - & - & - \\
            6 & - & - & - & - & - & - & 162.24 & 6.16 & 21.717 & 13.1879 & 31.80\% & 0.467 & - & - & - & - & - & - \\
            \bottomrule
    	\end{tabular}
    \end{adjustbox}
    \label{tab:powermodes-yolov3}
    \caption{Powermodes on Yolo v3}
    \footnotesize
\end{table*}

\textbf{Comparison matrix: all results together}.
This section is full of charts, and it's possible that even expert readers might get lost on them.
For the sake of readability, tables in Figures~\ref{TinyYOLOEfficiencyComp} and ~\ref{AlexNetEfficiencyComp} groups results that assess the throughput and power efficiency of the most interesting CNNs for each platform, namely Tiny-YOLO and AlexNet, at the best of their performance.
%
%
\begin{table}[t!]
	\centering
    \begin{adjustbox}{width=\columnwidth}
    	\begin{tabular}{c|c|c|c|c}
            \multirow{2}{*}{\textbf{SoC}} & \textbf{Accel. Freq.} & \textbf{Host Freq.} & \textbf{T.Put} & \textbf{Esys}\\
            & \textbf{[MHz]} & \textbf{[MHz]} & \textbf{[FPS]} & \textbf{[FPS/Watt]}\\ 
            \midrule
            Xavier & 624.7 & 499.2 & 164.2 & 54.093\\
            XU+    & 200.0 & 2400  & 22.68 & 19.782\\
            TX2    & 765.0 & 499.2 & 58.64 & 8.1613\\
            \bottomrule
    	\end{tabular}
    \end{adjustbox}
    \label{TinyYOLOEfficiencyComp}
    \caption{TinyYOLO Power Efficiency Comparison}
    \footnotesize
\end{table}
%
%
\begin{table}[t!]
	\centering
    \begin{adjustbox}{width=\columnwidth}
    	\begin{tabular}{c|c|c|c|c}
            \multirow{2}{*}{\textbf{SoC}} & \textbf{Accel. Freq.} & \textbf{Host Freq.} & \textbf{T.Put} & \textbf{Esys}\\
            & \textbf{[MHz]} & \textbf{[MHz]} & \textbf{[FPS]} & \textbf{[FPS/Watt]}\\ 
            \midrule
            Xavier & 675.7 & 192.0 & 162.5 & 194.59\\
            XU+    & 200.0 & 2400  & 61.55 & 120.77\\
            TX2    & 943.5 & 343.6 & 107.6 & 22.906\\
            \bottomrule
    	\end{tabular}
    \end{adjustbox}
    \label{AlexNetEfficiencyComp}
    \caption{AlexNet Power Efficiency Comparison}
    \footnotesize
\end{table}

%% file: experiments_2.tex
\section{Experimental Settings}
\label{sec:network-specific}

\subsection{Varying datasets}
%
%
%
\begin{table}[t!]
	\centering
    \begin{adjustbox}{width=\columnwidth}
    	\begin{tabular}{c|c|c|c|c|c}
            & \multirow{2}{*}{\textbf{Type}} & \multirow{2}{*}{\textbf{\#classes}} & \textbf{First layer} & \textbf{First layer} & \textbf{Validation test}\\
            &&&\textbf{(Yolo)}&\textbf{(Tiny-Yolo)}&\textbf{size}\\
            \midrule
            \textbf{COCO} & Generic & 80 & $608\times 608$ & $416\times 416$ & 1000\\
            \textbf{VOC} & Generic & 20 & $608\times 608$ & $416\times 416$ & 1000\\
            \textbf{Berkeley A} &\multirow{2}{*}{Automotive}&\multirow{2}{*}{3}&\multirow{2}{*}{$512\times 512$}&\multirow{2}{*}{-}&\multirow{2}{*}{1000}\\
            \textbf{(reduced)} &&&&&\\
            \textbf{Berkeley B} &\multirow{2}{*}{Automotive}&\multirow{2}{*}{10}&\multirow{2}{*}{$512\times 288$}&\multirow{2}{*}{-}&\multirow{2}{*}{1000}\\
            \textbf{(original)} &&&&&\\
            \bottomrule
    	\end{tabular}
    \end{adjustbox}
    \label{tab:datasetInfo}
    \caption{Data sets that we used in our exploration}
    \footnotesize
\end{table}
Table~\ref{tab:datasetInfo} shows the main features of the four datasets we employ in our analysis.
Datasets highly affect network precision, and (how we will now show) also performance.
Often, network engineers undergo a complete retraining of their network with small variation of the same dataset, e.g., removing unnecessary classes, to obtain highly-tuned optimal performance.
In our work we decided not to explore this dimension for all the networks and dataset, as to do so would make the design space explode.
We nonetheless analyzed a reduced version of ``Berkeley, with only three classes, hence, potentially faster and more accurate than the reference full version of the dataset.
%
%
\begin{table*}[t!]
	\centering
    \begin{adjustbox}{width=\textwidth}
    	\begin{tabular}{ll|cccccc|cccccc|cccccc}
            && \multicolumn{6}{c|}{\textbf{TX2}} & \multicolumn{6}{c|}{\textbf{Xavier}} & \multicolumn{6}{c}{\textbf{Nano}}\\
            \bottomrule
            \multirow{2}{*}{\textbf{Model}} & \multirow{2}{*}{\textbf{Training set}} & \textbf{E2E} & \multirow{2}{*}{\textbf{T.Put}} & \textbf{Pow} & \textbf{Pow} & \multirow{2}{*}{\textbf{mAP}} & \multirow{2}{*}{\textbf{P.eff}} & \textbf{E2E} & \multirow{2}{*}{\textbf{T.Put}} & \textbf{Pow} & \textbf{Pow} & \multirow{2}{*}{\textbf{mAP}} & \multirow{2}{*}{\textbf{P.eff}} & \textbf{E2E} & \multirow{2}{*}{\textbf{T.Put}} & \textbf{Pow} & \textbf{Pow} & \multirow{2}{*}{\textbf{mAP}} & \multirow{2}{*}{\textbf{P.eff}}\\

            & & \textbf{latency} & & \textbf{(Board)} & \textbf{(SoC)} & & & \textbf{latency} & & \textbf{(Board)} & \textbf{(SoC)} & & & \textbf{latency} & & \textbf{(Board)} & \textbf{(SoC)} & &     \\
            \midrule
            Yolo v2      & COCO     & 145.10 & 6.88     & 12.54  & 8.873  & 23.90\% & 0.775 & 62.07  & 16.11   & 37.354  & 26.98  & 23.90\% & 0.597 & 405.39 & 2.466 & 8.9303 & 6.7476 & 23.90\% & 0.365 \\
            Yolo v2      & VOC      & 70.10  & 14.3     & 10.678 & 7.011  & 58.82\% & 2.040 & 32.46  & 30.8    & 32.87   & 22.496 & 58.77\% & 1.369 & 175.69 & 5.691 & 9.5826 & 7.3999 & 58.77\% & 0.769 \\
            YOlo v2-Tiny & COCO     & 31.40  & 31.77    & 7.999  & 4.332  & 5.30\%  & 7.334 & 11.81  & 84.67   & 22.8162 & 12.442 & 5.30\%  & 6.805 & 60.84  & 16.43 & 5.94   & 3.7573 & 5.30\%  & 4.373 \\
            Yolo v3-Tiny & COCO     & 13.29  & 75.26    &10.309  & 4.116  & 6.48\%  & 18.287& 4.97   & 201.34  & 17.2258 & 6.4750 & 6.48\%  & 31.095& 42.01  & 23.80 & 7.97   & 2.8877 & 6.54\%  & 8.243 \\
            Yolo v3      & VOC      & 147.98 & 6.76     & 14.712 & 4.116  & 78.35\% & 1.642 & 73.22  & 13.658  & 32.255  & 6.475  & 78.35\% & 2.109 & -      & -     & -      & -      & -       & -     \\
            Yolo v3      & COCO     & 291.48 & 3.43     &15.105  & 11.438 & 31.80\% & 0.300 & 106.59 & 9.38    & 38.399  & 28.025 & 31.80\% & 0.335 & -      & -     & -      & -      & -       & -     \\
            Yolo v3      & Berkeley & 129.90 & 7.69     & 14.174 & 10.507 & 38.50\% & 0.732 & 61.52  & 16.25   & 41.515  & 31.141 & 38.57\% & 0.522 & -      & -     & -      & -      & -       & -     \\
            \bottomrule
    	\end{tabular}
    \end{adjustbox}
    \label{tab:datasets1}
    \caption{Floating point versions of Yolo v3, varying  datasets on GPGPU}
    \footnotesize
\end{table*}
%
%
\begin{table*}[t!]
	\centering
    \begin{adjustbox}{width=\textwidth}
    	\begin{tabular}{ll|cccccc|cccccc}
            && \multicolumn{6}{c|}{\textbf{Ultra96}} & \multicolumn{6}{c}{\textbf{ZCU102}} \\
            \bottomrule
            \multirow{2}{*}{\textbf{Model}} & \multirow{2}{*}{\textbf{Training set}} & \textbf{E2E} & \multirow{2}{*}{\textbf{T.Put}} & \textbf{Pow} & \textbf{Pow} & \multirow{2}{*}{\textbf{mAP}} & \multirow{2}{*}{\textbf{P.eff}} & \textbf{E2E} & \multirow{2}{*}{\textbf{T.Put}} & \textbf{Pow} & \textbf{Pow} & \multirow{2}{*}{\textbf{mAP}} & \multirow{2}{*}{\textbf{P.eff}}\\

            & & \textbf{latency} & & \textbf{(Board)} & \textbf{(SoC)} & & & \textbf{latency} & & \textbf{(Board)} & \textbf{(SoC)} \\
            \midrule
            Yolo v2 & COCO & - & - & - & - & - & - & - & - & - & - & - & - \\
            Yolo v2 & VOC  & 85.956 & 11.5 & 9.18 & 2.604 & 63.78\% & 4.416 & 40.064 & 24.96 & 37.824 & 5.46 & 64.56\% & 4.571 \\
            Yolo v2-Tiny & COCO & - & - & - & - & - & - & - & - & - & - & - & - \\
            Yolo v3-Tiny & COCO & 83.542 & 11.97 & 7.836 & 1.26 & 4.60\% & 9.500 & 71.994 & 13.89 & 34.44 & 2.076 & 4.60\% & 6.691 \\
            Yolo v3 & VOC  & 150.82 & 6.63 & 9.912 & 3.336 & 73.17\% & 1.987 & 69.444 & 14.4 & 38.364 & 6.00 & 75.07\% & 2.400 \\
            Yolo v3 & COCO  & 950.57 & 1.052 & 8.088 & 1.512 & 31.30\% & 0.696 & 319.3 & 3.1318 & 34.656 & 2.292 & 31.30\% & 1.366 \\
            Yolo v3 & Berkeley (A)  & 22.492 & 44.46 & 9.576 & 3.00 & 19.44\% & 14.820 & 11.76 & 85.03 & 36.552 & 4.188 & 19.44\% & 20.303 \\
            Yolo v3 & Berkeley (B)  & 157.9 & 6.333 & 9.912 & 3.336 & 55.20\% & 1.898 & 72.463 & 13.80 & 37.848 & 5.484 & 55.20\% & 2.516 \\
            \bottomrule
    	\end{tabular}
    \end{adjustbox}
    \label{tab:datasets2}
    \caption{Quantized versions of Yolo v3, varying datasets on FPGA}
    \footnotesize
\end{table*}
Table(s)~\ref{tab:datasets1} and ~\ref{tab:datasets2} show how target datasets perform on selected networks and platforms. 
Looking at mAP, we see it is in line with what expected and declared in the reference papers and guides for each network and dataset.
Also, there is a slight difference in terms of precision between GPGPUs and FPGAs, due to low-level optimizations such as data types (FPGAs use INT8, while GPGPUs use different optimization for different networks, most of which are un-disclosed in the production environment).

One interesting thing to note is that different networks and datasets affect system performance (throughput and E2E latency), which in turn affects power consumption.

\textbf{Observation}: \textit{there is an (in)direct correlation between the dataset and the power consumption.}

Also here, we can note how FPGA-based platforms are at least twice power efficient as GPGPUs.
Among GPGPUs, strangely, TX2 outperforms Xavier in power efficiency, and also Nano offer similar ($\approx 4$ vs. $\approx$ 6) numbers, while consuming $ \approx 1/4 $ of power.

\subsection{Quantization (FPGA only)}
%
%
\begin{table*}[t!]
	\centering
    \begin{adjustbox}{width=\textwidth}
    	\begin{tabular}{l|cccccc|cccccc}
            & \multicolumn{6}{c|}{\textbf{Ultra96}} & \multicolumn{6}{c}{\textbf{ZCU102}} \\
            \bottomrule
            \multirow{2}{*}{\textbf{Quantization}} & \textbf{E2E} & \multirow{2}{*}{\textbf{T.Put}} & \textbf{Pow} & \textbf{Pow} & \multirow{2}{*}{\textbf{mAP}} & \multirow{2}{*}{\textbf{P.eff}} & \textbf{E2E} & \multirow{2}{*}{\textbf{T.Put}} & \textbf{Pow} & \textbf{Pow} & \multirow{2}{*}{\textbf{mAP}} & \multirow{2}{*}{\textbf{P.eff}}\\

            & \textbf{latency} & & \textbf{(Board)} & \textbf{(SoC)} & & & \textbf{latency} & & \textbf{(Board)} & \textbf{(SoC)} \\
            \midrule
            INT8            & 86.956 & 11.5 & 9.180  & 2.604 & 63.78\% & 4.41628 & 40.064 & 24.96 & 37.824 & 5.46  & 64.56\% & 4.57140 \\
            INT8 Pruned 22g & 31.847 & 31.4 & 9.744 & 3.168 & 62.94\% & 9.91161 & 19.109 & 52.33 & 37.092 & 4.728 & 63.10\% & 11.0681\\
            INT8 Pruned 24g & 28.116 & 35.56 & 9.684 & 3.108 & 61.41\% & 11.4436 & 17.143 & 58.33 & 36.756 & 4.392 & 61.96\% & 13.2809\\
            INT8 Pruned 26g & 24.473 & 40.86 & 9.702 & 3.126 & 61.59\% & 13.0710 & 15.213 & 65.73 & 36.984 & 4.62 & 61.25\% & 14.2272\\
            \bottomrule
    	\end{tabular}
    \end{adjustbox}
    \label{tab:quantization}
    \caption{Quantized and pruned version of the Yolo v2 provided with Xilinx SDK}
    \footnotesize
\end{table*}
Quantization is one of the most significant optimization that can be employed on artificial neural networks, as they affect both performance and precision. The aim of quantization phase is to employ fixed point integers for network's weights and activation to better exploit the characteristics of the underlying hardware, without heavily impact on average precision \cite{quantization}
Table(s)~\ref{tab:datasets2} shows how a selected network and dataset  performs under quantization.
In this work we used INT8 quantization only on FPGA-based systems, because GPGPUs are highly efficient on performing floating-point computing, so we do not expect big improvement in terms of mAP/efficiency on these latter architectures.
We show results for a single network and dataset (namely, Yolo v2 trained with VOC), as we want to show the variation of performance compared to baseline (in this case, INT8), however, similar results hold also for other configuration.

\subsection{Pruning (FPGA only)}
Pruning a neural network means reducing its size and complexity by removing parameters and activations that "do not play a big role" inside the whole network~\cite{pruning}.
Also in this case we did not exploit this technique on NVIDIA GPGPUs, because their inference engine framework, called \textit{TensorRT}, does not allow network pruning.
We show results in Table~\ref{tab:quantization}.
The 'g' of names in the first column ('22g', '24g', '26g'), is used by Xilinx to label the declared performance in GFLOPs of the network. Intuitively, the higher the number, the higher the throughput, and this is reflected in our tables.
Results are available only for FPGA platforms, as such a fine-tuning of the network is not possible in the NVIDIA boards.
We see how pruning and quantization affects performance, but this does not reflect in higher power consumption, nor harnesses network precision.

\textbf{Observation}: \textit{This apparently counter-intuitive result is due to low-level network optimization (remember that FPGA-based system enable engineers coding neurons nearly in hardware), and enables higher power efficiency almost ``for free'' at the same precision.}
\subsection{Tiny networks}
%
%
\begin{table}[t!]
	\centering
    \begin{adjustbox}{width=\columnwidth}
    	\begin{tabular}{cl|cccccc}
            &&\textbf{E2E latency} & \textbf{T.Put} & \textbf{Pow (Board)} & \textbf{Pow (SoC)} & \textbf{mAP} & \textbf{P.eff}\\
            \midrule
            \multirow{6}{*}{Full Yolo V3} & TX2     & 291.48 & 3.43   & 15.11 & 11.44 & 31.80\% & 0.30\\
                                          & Xavier  & 106.59 & 9.38   & 38.40 & 28.03 & 31.80\% & 0.33\\
                                          & NANO    & -      & -      & -     & -     & -       & -   \\
                                          & Ultra96 & 950.57 & 1.052  & 8.09  & 1.51  & 31.30\% & 0.70\\
                                          & ZCU102  & 319.00 & 3.1318 & 34.66 & 2.29  & 31.30\% & 1.37\\
            \midrule
            \multirow{6}{*}{Full Yolo V2} & TX2     & 145.10 & 6.88   & 12.54 & 8.87  & 23.90\% & 0.78\\
                                          & Xavier  & 62.07  & 16.11  & 37.35 & 26.98 & 23.90\% & 0.60\\
                                          & NANO    & 405.39 & 2.466  & 8.93  & 6.75  & 23.90\% & 0.37\\
                                          & Ultra96 & 86.96  & 11.5   & 9.18  & 2.60  & 63.78\% & 4.42\\
                                          & ZCU102  & 400.64 & 24.96  & 37.82 & 5.46  & 64.56\% & 4.57\\
            \midrule
            \multirow{6}{*}{Tiny Yolo V2} & TX2     & 31.40  & 31.77  & 8.00  & 4.33  & 5.30\%  & 7.33\\
                                          & Xavier  & 11.81  & 84.67  & 22.82 & 12.44 & 5.30\%  & 6.81\\
                                          & NANO    & 60.84  & 16.43  & 5.94  & 3.76  & 5.30\%  & 4.37\\
                                          & Ultra96 & -      & -      & -     & -     & -       &    -\\
                                          & ZCU102  & -      & -      & -     & -     & -       &    -\\
            \midrule
            \multirow{6}{*}{Tiny Yolo V3} & TX2     & 13.29  & 75.262 & 10.31 & 4.12  & 6.48\%  & 18.29\\
                                          & Xavier  & 4.97   & 201.34 & 17.23 & 6.48  & 6.48\%  & 31.10\\
                                          & NANO    & 42.01  & 23.804 & 7.97  & 2.89  & 6.54\%  &  8.24\\
                                          & Ultra96 & 835.42 & 11.97  & 7.84  & 1.26  & 4.60\%  &  9.50\\
                                          & ZCU102  & 719.94 & 13.89  & 34.44 & 2.08  & 4.60\%  &  6.69\\
            \bottomrule
    	\end{tabular}
    \end{adjustbox}
    \label{tab:tiny-coco}
    \caption{Tiny version of the YOLO v2 and v3 on reference platforms}
    \footnotesize
\end{table}
Small network variants are extremely suitable for low-end, resource-constrained platforms~\footnote{Numbers shown refer to experiments with the COCO dataset}.
For instance, the NVIDIA NANO cannot support a full YOLO v3 network
On the other hand, it might not be convenient  to deploy a reduced network version on some platforms.
This explains the empty cells in Table~\ref{tab:tiny-coco}, where we show how different network variants perform on different platforms.

\textbf{Observation}: \textit{Intuitively, while smaller platforms achieve highest power efficiency in running smaller networks, it might not always be worth, such as for instance in the case of Tiny-Yolo on Ultra96, that exhibit very poor mAP of $ \approx 4.6\% $.}

\subsection{Batching/Threading}
\label{subsec:batching-threading}
The most effective/important optimization we employ consists of loading multiple images into the accelerator, this technique allows to improve the utilization  of the accelerator. 
For example, each layer of the network will have some amount of overhead and  synchronization required to compute forward inference.
By computing more results in parallel, this overhead is paid off more efficiently.
In NVIDIA platforms, this feature is called \textit{batching}, and it is typically performed by a single application thread that ``packs'' images in a \textit{batch} transfer to the GPGPU accelerator.
On the other hand, the FPGA-based platforms that we target provided \textit{multi-threaded} version of the host-side application, which is capable of delivering multiple images to the DNN IP exploiting the multicore host and multiple DNN cores.
In our experiments, batching/threading does not affect network precision (mAP), hence, we don't report it.

	
\textbf{Detection networks}.
%
%
%
\begin{table*}[t!]
	\centering
    \begin{adjustbox}{width=\textwidth}
    	\begin{tabular}{lc|cccccc|cccc}
            \multirow{2}{*}{\textbf{Network}} & \textbf{Batch size} & \multicolumn{2}{c}{\textbf{TX2}} & \multicolumn{2}{c}{\textbf{Xavier}} & \multicolumn{2}{c|}{\textbf{Nano}} & \multicolumn{2}{c}{\textbf{Ultra96}} & \multicolumn{2}{c}{\textbf{ZCU102}}\\
            & \textbf{/Nthreads} & \textbf{E2E latency} & \textbf{T.Put} & \textbf{E2E latency} & \textbf{T.Put} & \textbf{E2E latency} & \textbf{T.Put} & \textbf{E2E latency} & \textbf{T.Put} & \textbf{E2E latency} & \textbf{T.Put}\\
            \midrule
            Full Yolo & 1 & 303.41  & 3.296 & 153.38 & 6.520 & - & - & 950.57 & 1.052 & 319.30 & 3.132\\
            Full Yolo & 2 & 604.76  & 1.654 & 293.42 & 3.408 & - & - & 950.57 & 1.959 & 319.30 & 6.132\\
            Full Yolo & 4 & 1215.53 & 0.823 & 583.09 & 1.715 & - & - & 950.57 & 3.007 & 319.30 & 11.459\\
            Full Yolo & 8 & 2397.03 & 0.417 & 1156.41 & 0.865 & - & - & 950.57 & 3.051 & 319.30 & 15.662\\
            Full Yolo & 16 & 4826.71 & 0.207 & 2362.29 & 0.423 & - & - & 950.57 & 3.068 & 319.30 & 15.586\\
            \midrule
            Yolo Tiny & 1 & 13.29  & 75.262 & 4.97 & 201.341 & 42.01 & 23.804 & 83.54 & 11.97  & 71.99 & 13.890\\
            Yolo Tiny & 2 & 26.38  & 37.911 & 9.57 & 104.520 & 82.74 & 12.086 & 83.54 & 21.189 & 71.99 & 27.530\\
            Yolo Tiny & 4 & 52.54 & 19.032 & 18.79 & 53.230 & 166.37 & 6.011 & 83.54 & 36.548 & 71.99 & 52.930\\
            Yolo Tiny & 8 & 106.28 & 9.410 & 37.44 & 26.708 & 338.02 & 2.958 & 83.54 & 39.219 & 71.99 & 59.410\\
            Yolo Tiny & 16 & 211.17 & 4.736 & 74.50 & 13.422 & 678.90 & 1.473 & 83.54 & 35.96 & 71.99 & 59.860\\
            \bottomrule
    	\end{tabular}
    \end{adjustbox}
    \label{tab:batching-threading}
    \caption{Effect of batching/multi-threading on Full and Tiny version of the Yolo v3}
    \footnotesize
\end{table*}
We carred on experiments on the most heavyweight and powerful network (YOLO v3 trained with COCO), and results are reported in Table~\ref{tab:batching-threading}
Surprisingly, they show how, for the same network, performance do not vary much, and this is because of the size of network itself and of input frames (80 classes, frames are 608x608, see Table~\ref{tab:datasetInfo}).
On the other hand, on FPGA-based platforms we see interesting results.
High parallelism in Ultra96 only slightly increases performance, either for Full or Tiny Yolo V3.
The reason is that, the board is already overloaded by computation on this heavyweight network.
On the other hand, because of the higher number of DNN cores instantiated, ZCU102 can still keep the scaling pace up to 4-8 thread (resp. for the two datasets), but then it also ``surrends'' to the heavy workload.

One consideration we must draw on the different approaches to parallelism offered by the two platform families.
While, apparently, the effect is the same (we increase accelerator utilization by offloading multiple data frames), as shown, the two approaches are deeply different.

\begin{figure}[t!]
	\includegraphics[width=0.8\columnwidth]{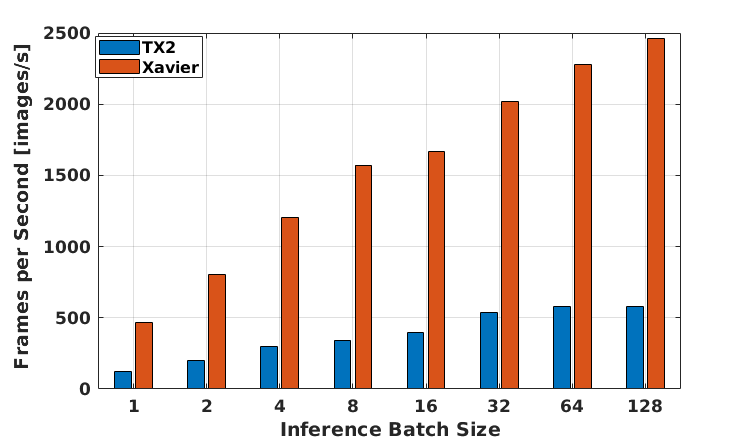}
	\centering
	\caption{Framerate achieved by AlexNet on TX2 and Xavier, quantized with 16-bit floating point numbers, varying the \textit{batch size} parameter}
	\label{BatchSizeFPS}
\end{figure}

\begin{figure}[t!]
	\includegraphics[width=0.8\columnwidth]{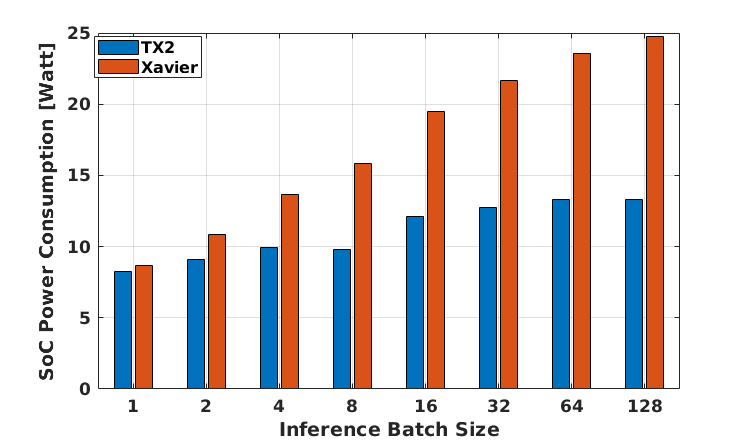}
	\centering
	\caption{Power consumption of AlexNet on TX2 and Xavier, quantized with 16-bit floating point numbers, varying the \textit{batch size} parameter}
	\label{BatchSizePower}
\end{figure}

\textbf{Classification networks}.
In Figure \ref{BatchSizeFPS} and \ref{BatchSizePower} we can see the effect of batching multiple input images on GPGPU system, in terms of FPS and power consumption.
From the graphs shown, we can see that through batching, TX2 and Xavier were able to reach about $500$ and $2500$ FPS respectively in $13$ and $25$ Watts, and performance in Xavier could potentially scale more with the batch size.

\textbf{Observation}: \textit{GPGPU batching might be ineffective with big detection networks, because the host system might not  be able to process and pack all camera frames in real-time. Threading, instead, might play a role, where available, because it enables multiple concurrent stream of data.}

Finally, few very important final consideration follow.

\textbf{Observation}: \textit{in none of our tests we caused a bottleneck neither in system bus nor in system memory, when transferring frames back and forth the accelerator. This seems to indicate that the complex memory/bus hierarchy of modern platforms is such powerful, that, at least for the considered workloads, its limits are still far away to be reached.}

\textbf{Reasoning on results}.
In GPGPU systems, the application code must collect a number of data frame (i.e., images from multiple cameras), before being able to pack and offload it to the accelerator.
Especially, in NVIDIA systems, the NN package and inference engine drivers give poor freedom to the system designer to access camera in parallel from within the same application.
This aspect could certainly can be improved, if the system were not proprietary and closed.
A possible enhancement for reconfigurable platforms, and especially Xilinx's SoMs, is that it is possible for engineers to direct the video stream across the FPGA to get through the DRAM banks.
This point is really interesting, as it potentially enables on-the-flight detection on the video stream without involving the host cores, and deserves more investigation from system engineers and experts of hardware design.

%% file: conclusions.tex
\section{Conclusions}
\label{sec:conclusions}

In this work, we evaluated the power-efficiency and performance of image classification and objects detection Convolutional Neural Networks, on embedded platforms representative of next generation automotive domain controllers.
Our results assess higher peek performance, in terms of frame-per-second on GPGPU-based accelerators compared to the reconfigurable logics.
Regarding the AlexNet network, with the batch size parameter set at $128$, the new NVIDIA Xavier is resulted faster than TX2 of about $5\times$, with an increase factor of $1.92\times$ in power consumption.
We experienced comparable performance for the object detection networks, due to their complexity and size, while for classification network, GPGPUs still outperform (non-optimized) FPGAs by at least one order of magnitude.
On the other hand, FPGAs are better from the power consumption and power-efficiency viewpoint.
Figures \ref{AlexNetEfficiencyComp} and \ref{TinyYOLOEfficiencyComp} show what are probably the most interesting (and useful) outcomes of our effort, a comparison of the two main CNN models, namely AlexNet and Tiny-YOLO, in terms of power-efficiency (Formula \ref{eff}).
They show the highest power-efficiency of XU+, because the power consumption due to the FPGA fabric is very low, which was somehow expected.

We also explored both platform optimizations, namely power scaling/profiles, batching and threading, and network-specific features such as quantization and multiple datasets, assessing not only the performance/power dimension, but also how they do affect network precision.
Interesting results act as guidelines for engineers, and as starting point for future works, when new platforms and networks come to light.

%% file: appendix_methodology.tex
\section{Methodology for power measurement}
\begin{figure}[h!]
	\includegraphics[width=.9\columnwidth]{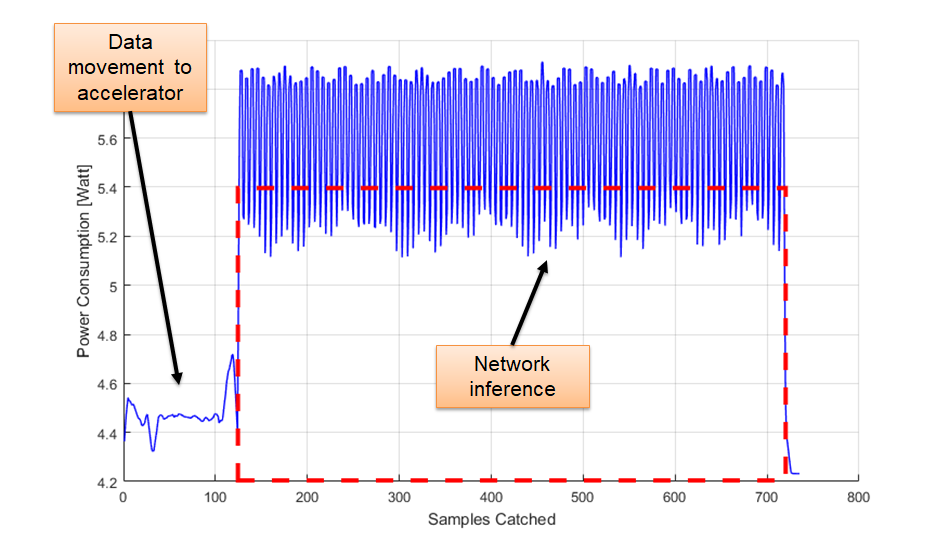}
	\centering
	\caption{approximation of the power consumption during the inference phase}
	\label{inference}
\end{figure}
Accurately measuring the power consumption is a challenging task, because the different hardware platforms considered in this work have different set of sensors and primitives.

The methodology we used for the \textbf{Zynq UltraScale+} is based on measuring the current that flows from the power supply into the Zynq board, because there are power sensors/counters that are not available on that board.
This current flow was measured through a multimeter (configured as Amperometer) connected in series as the \textit{VDD} and \textit{GND} cables of the board power supply.\\
The total current absorption of the development board was measured with the precision of 40 samples-per-second, that is the highest for the multimeter that we used.
Based on the samples caught by the amperometer, the dynamic power consumption
%
%
has been approximated by multiplying the mean current absorbed and the voltage supply with the formula:
\begin{equation}
Pow_{(board)} = V_{DD}\times I_{mean}
\label{pboard}
\end{equation}
Then we derived $P_{inf}$ a power measure obtained by algebraic subtraction of the power computed with the previous formula and the power measured when the \textit{System-on-Chip (SoC)} is in idle: 

%
\begin{equation}
Pow_{(SoC)} = Pow_{(board)} - Pow_{(idle)}
\label{pinf}
\end{equation}
This approximate metric roughly captures the power actually consumed in the inference process.
Subsequently we approximate the power-efficiency of the system, combining FPS and power dissipation as in the following formula:

\begin{equation}
\begin{minipage}{.40\linewidth}
	\centering
$ P.eff= \frac{FPS}{P_{mean}}\qquad $
\end{minipage}
\begin{minipage}{.40\linewidth}
	\centering
$ \left[ \frac{images\,/\,s}{Watt} = {images}\,/\,J \right] $
\end{minipage}
\label{eff}
\end{equation}

For the power consumption of the FPGA, we compute the power consumption with the formulas \ref{pboard} and \ref{pinf}.
On the other hand, \textbf{NVIDIA Tegras} boards have on-board sensors (and related APIs directly accessible from software), hence for these platforms we directly access the power consumption using sensors.

Finally for NVIDIA TX2 and Jetson AGX Xavier, like the previous case, we used a sensor-based approach, through the \textit{INA3221} sensor\footnote{The \texttt{sysfs} nodes to read power on Jetson are located in \texttt{/sys/bus/i2c/drivers/ina3221x}.}.
This sensor has three channels that can monitor respectively the \textit{VDD\_IN}, \textit{VDD\_CPU} and \textit{VDD\_DDR} power rails, through an \textit{I2C} connection.

For these experiments we need to monitor the \textit{VDD\_IN} rail, because is the rail that provides energy to the SoC itself.
In Figure \ref{inference} we can see the typical trend in power consumption, that for example we can compute trough the Formula \ref{pboard} in which we can recognize the \textit{image loading phase} (on the left in the graph) and the inference phase (on the right), that consists of a sequence of power peaks.
The dashed rectangle shows an approximation of the total energy needed during the inference phase.

It is important to note that we also run a reduced set of experiments by applying the methodology we used for the XU+ board i.e., physically cutting rails) also to the NVIDIA boards, and measured power was comparable to the ones from sensors.
This proves that the approach of physically acting on the board is accurate enough for our purposes (besides the fact that it damages the board itself).

%% file: appendix_frameworks.tex
\subsection{Evaluated Frameworks}
\label{subsec:appendix-frameworks}
Inference engines play a big role in maximising the performance of a network, on every platform, and for this reason they usually come with the SDK by platform provider.
In our exploration, we employed the reference framework for every specific network and platform.
This is the typical choice of system engineers, because it represents the best tradeoff among the metrics, and it's supported by the community and platform provider.

\textbf{TensorRT} is an optimized Deep Neural Network library for accelerating the inference phase, to exploit TensorRT we used tkDNN~\footnote{https://github.com/ceccocats/tkDNN}, a framework written in C++ and CUDA.
tkDNN is optimized to execute mainly on embedded SoC like for example the NVIDIA Tegra family, rather than GPU-based servers.
We also used \textit{TensorRT} directly, (through the \textit{trtexec} tool), for testing AlexNet on TX2 and Xavier.

\textbf{DPU ecosystem} the Xilinx DPU is an optimized engine to accelerate DNNs on top of Xilinx's SoCs.
DPU is released as IP-core with Xilinx SDK, and it can be easily inserted in a custom hardware design.
The DPU operation requires the application processing unit (APU) to service interrupts to coordinate data transfer.
DNNDK is a framework provided by Xilinx in the Ultrascale SDK, that allows to quantize and compile a deep learning model for a target DPU architecture.
A machine learning engineer can define a custom neural network model with some of the most popular engines, like Caffe or Tensorflow and after the training phase it is possible to convert the model to fixed point and deploy it to a DPU engine.
At the moment the DNNDK, framework can quantize a NN with an arbitrary fixed point precision, but the DPUv2 in this current release supports only the INT8 quantization, for these reason all the networks quantized with this framework have this type of quantization.

\textbf{Other Frameworks} tested in this work are Caffe, a deep learning framework developed by the Berkeley Vision and Learning Center (BVLC) and proposed in \cite{Caffe}.
Caffe is a standard for implementing CNNs and it is well integrated in several others frameworks and libraries for embedded accelerators.
xfDNN is a framework based on Caffe for accelerating deep neural networks on Xilinx UltraScale+ MPSoCs. This framework supports 16-bit integer data type.
It became an opensource project called CHaiDNN \cite{CHaiDNN}.

PipeCNN is another CNN framework, written in OpenCL and proposed by D. Wang et al. in \cite{PipeCNN} that we used for implementing AlexNet on Xilinx UltraScale+.
This framework is based on \textit{pipes}, a new feature introduced in OpenCL 2.x.